\documentclass[journal,comsoc, 11pt]{IEEEtran}

\usepackage[T1]{fontenc}

\usepackage[crop=pdfcrop,cleanup={.tex, .dvi, .ps, .pdf, .log, .aux, .bbl, .out}]{pstool}

\newcommand{\VG}[1]{{
	\let\theta\uptheta
	\let\mu\upmu
	\let\beta\upbeta
	\let\gamma\upgamma
	\bm{#1}
	}}

\begin{document}
\title{Quickest Detection and Forecast of Pandemic Outbreaks: Analysis of COVID-19 Waves}

\author{
Giovanni~Soldi, Nicola Forti, Domenico~Gaglione, Paolo~Braca,~\IEEEmembership{Senior Member,~IEEE,} \\
Leonardo M. Millefiori,~\IEEEmembership{Member,~IEEE,}
Stefano Marano,~\IEEEmembership{Senior Member,~IEEE,} \\
Peter K. Willett,~\IEEEmembership{Fellow,~IEEE,}
Krishna R. Pattipati~\IEEEmembership{Life Fellow,~IEEE}
\vspace{-10mm}
\thanks{G. Soldi, N. Forti, D. Gaglione, P. Braca, and L. M. Millefiori are with the NATO Centre for Maritime Research and Experimentation (CMRE). Their work was supported by the Data Knowledge and Operational Effectiveness (DKOE) program, sponsored by the NATO Allied Command Transformation (ACT). S. Marano is with University of Salerno. P. Willett and K. Pattipati  are with the University of Connecticut. The work of P. Willett was supported in part by AFOSR under Contract FA9500-18-1-0463. The work of K. Pattipati was supported in part by the U.S. ONR, in part by the U.S. NRL under Grant N00014-18-1-1238 and Grant N00173-16-1-G905, and in part by the NASA's Space Technology Research Grants Program under Grant 80NSSC19K1076.}
}

\maketitle

\begin{abstract}
The COVID-19 pandemic has, worldwide and up to December 2020, caused over 1.7 million deaths, and put the world's most advanced healthcare systems under heavy stress.  
In many countries, drastic restrictive measures adopted by political authorities, such as national lockdowns, have not prevented the outbreak of
new pandemic's
waves.
In this article, we propose an integrated detection-estimation-forecasting framework that, using
publicly available data,
is designed to: (i) learn relevant features of the
pandemic (e.g., the infection rate); (ii) detect as quickly
as possible the onset (or the termination) of an exponential growth of the contagion; and (iii) reliably forecast the 
pandemic evolution.
The proposed solution is
validated by analyzing
the COVID-19 second and third
waves in the USA.
\end{abstract}


\IEEEpeerreviewmaketitle

\vspace{-4mm}
\section{Introduction}

\bstctlcite{IEEEexample:BSTcontrol}

\IEEEPARstart{O}{n} March 11, 2020, the World Health Organization (WHO) declared the COVID-19 disease a pandemic.
Since then, many governments,
hampered by the lack of an effective cure, decided to undertake extraordinary social measures, such as travel bans, closure of schools, universities, shops, factories and even national lockdowns, causing disruptive changes in social behavior, global mobility patterns, and the economies, see e.g.~\cite{millefiori2020covid19}.
These measures resulted in effectively reducing the infection rate and slowing the spread of the pandemic~\cite{DehZieSpiPau:J20}, thereby bringing the number of cases under control and relieving the pressure on the intensive care units.
However, because of the premature relaxation of these measures, new waves of COVID-19 cases are rampant in many countries around the world.
Despite the experience of the first wave, many governments have failed to detect these
new exponential growth patterns early,
and, consequently, either have acted too late or have applied
light and ineffective countermeasures.
This suggests that it is
of paramount importance to develop advanced models and algorithms that are able to detect the onset of an exponential growth phase as quickly as possible, and to forecast the incipient evolution of the infection in order to provide local and governmental authorities with enhanced real-time decision support.

Leveraging our knowledge in quickest detection techniques~\cite{BassevilleBook,poor-book-quickest}, adaptive Bayesian filtering and target tracking~\cite{bar2011tracking},
we propose a framework that, based on
data provided on a daily basis by authorities (e.g., number of
infected and recovered), is able to \textit{i)} learn relevant features of the
pandemic, e.g., the infection rate, \textit{ii)} detect as quickly as possible the {passages} from {(or back to)} a controlled regime, i.e., a phase during which the number of new cases is under control, to {(from)} a critical one, i.e., characterized by an exponential growth in the number of infected, and \textit{iii)}
reliably
forecast the
pandemic evolution. We exploit recently developed tools for quickest detection of COVID-19 pandemic onset, learning of its peculiar features, and forecasting its evolution.
The quickest detection task relies on a
method recently presented in~\cite{braca2020quickest_1} and \cite{braca2020quickest_2}, that is
a version of the celebrated Page's CUSUM test~\cite{BassevilleBook,poor-book-quickest} specifically tailored to non-stationary pandemic data.
This method, called the mean agnostic sequential test (MAST), is able to detect the onset of an exponential
pandemic growth by properly 
trading-off
the delay in intervention and the risk of incorrectly declaring an outbreak. {The MAST, properly adjusted (inverting the roles of the hypotheses), is also able to detect the termination of a pandemic wave as well.}
As for the learning and forecasting tasks, 
epidemiological compartmental models, such as the SIR and SEIR models (cf. Section~\ref{sec:epidemic_modeling}), are used to describe the
pandemic evolution.
Model parameters, such as infection
and recovery rates, are
considered
time-varying, and are
learned together with the posterior probability distributions of the main epidemiological quantities, e.g., the numbers of infected and recovered individuals; see details in~\cite{GagBraMilSolForMar:J20}.

Key to the accuracy of the forecast is to know which recent data apply to it; that is, we need the change-points between controlled and critical regimes.
For this reason, in this paper, we combine the quickest detection approach with the Bayesian forecast to develop an integrated detection-forecast framework. In particular, detecting the beginning and the termination of a pandemic wave through MAST enables a more reliable infection rate estimation to be adopted in accurate forecast of pandemic evolution up to several weeks after the detection.
The comprehensive set of tools provided by the proposed framework might assist the authorities in evaluating the implementation of pandemic countermeasures. 
The
effectiveness of the proposed framework is assessed by
{detecting the onsets
and terminations of the second and third waves} of COVID-19 in the USA, starting from May 1, 2020, and forecasting the evolution of the contagion up to December 13, 2020. 

The remainder of this paper is organized as follows. Section~\ref{sec:epidemic_modeling}
presents the most used compartmental models for epidemiological modeling. 
Section~\ref{sec:proposed_method}
describes the proposed framework
that includes the
Bayesian learning of the model parameters
the quickest detection of an exponential growth, 
and the forecasting of the
contagion.
Data analysis of the second and third waves of COVID-19 in the USA is presented in Section~\ref{sec:analysis_second_wave}, and
concluding remarks are provided in Section~\ref{sec:conclusion}.

\vspace{-4mm}

\section{Epidemiological Modeling}
\label{sec:epidemic_modeling}

\begin{figure*}
\centering
\psfragfig[width=.96\textwidth]{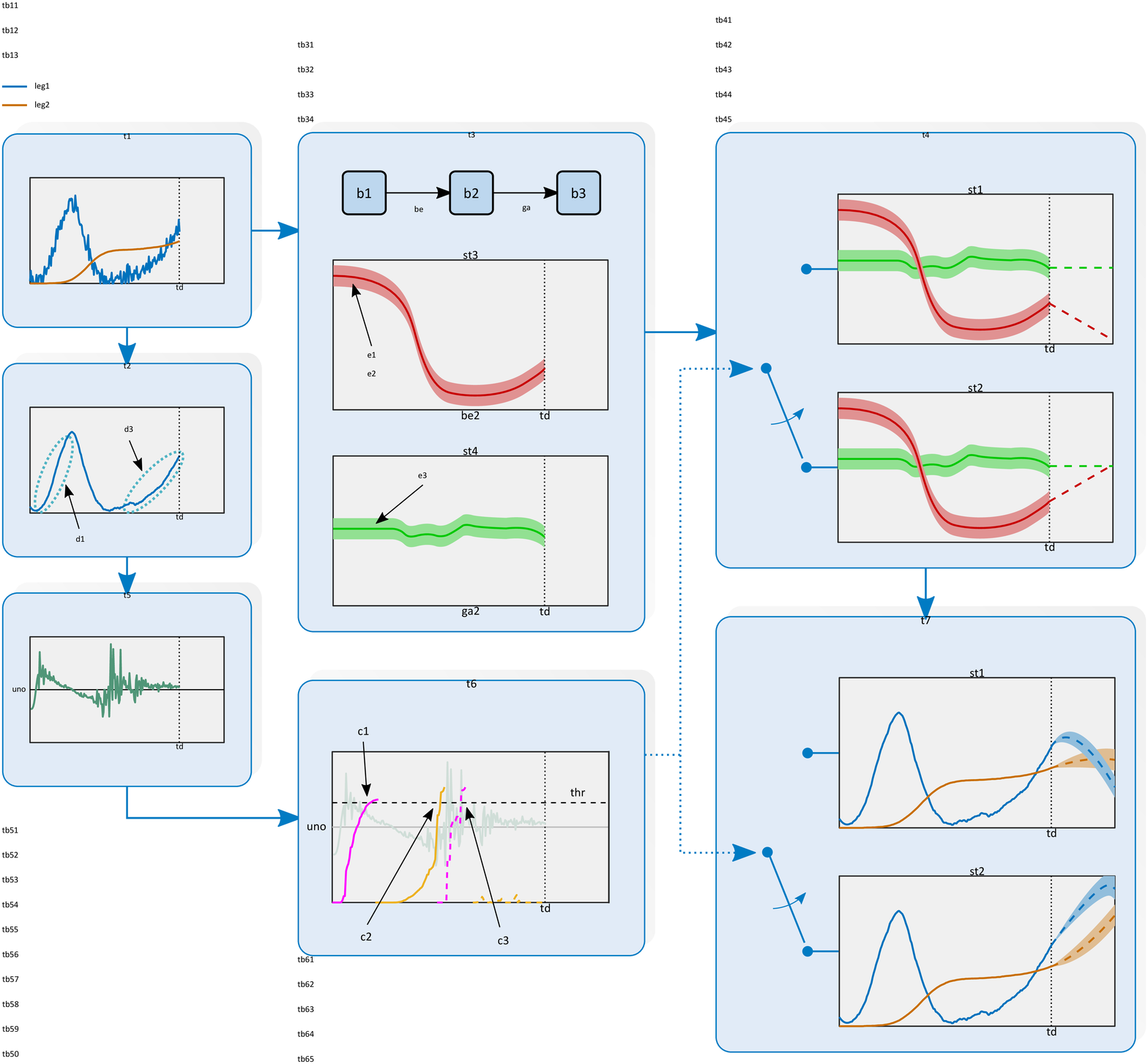}
\vspace{1mm}
\caption{
Notional sketch of the proposed decision-directed-forecasting framework.
}
\label{fig:flowchart}
\vspace{-5mm}
\end{figure*}

Compartmental epidemiological models assume that a given population is partitioned into a predefined number of compartments (population subgroups), where each compartment represents
a pandemic state that an individual can occupy.
The SIR model~
\cite{BjoSheLkrAlt:J20}
accounts for three 
compartments, specifically, susceptible (S), infected (I), and recovered (R) individuals. A susceptible individual can contract the virus at a fixed constant ``infection'' rate, denoting the rate at which the individual comes in contact with an infected individual. If infected, an individual develops the disease and is transitioned to the infected compartment.
Finally, an infected individual 
recovers or passes away at a constant ``recovery'' rate, thereby moving to the recovered compartment. Recovered people are considered permanently immune.

Over the years, more complex extensions of the SIR model have been developed. 
For example,
the SEIR model
assumes that a susceptible individual does not develop the disease symptoms immediately, but only after an incubation period 
of a certain duration (in the COVID-19 case, this duration ranges from 3 to 15 days, with a median value of 
5.2 days). Therefore, susceptible individuals
go through an 
exposed (E) compartment before developing evident
symptoms, and, eventually, move to the infected compartment.
In the SEIRQ model~\cite{HuQiaJunXiaWeiZhi:J20},
an extra compartment is added for individuals 
who have contracted the virus and are quarantined (Q).
A further extension is represented by the
generalized SEIR (GSEIR) \cite{PenYanZhaZhuHon:J20}, that includes three more compartments, i.e., insusceptible, quarantined, and death.
The SIR-X model~\cite{MaiBenBro:K20} takes 
into account restrictive measures, such as closure of schools and shops, or complete lockdown, by 
removing susceptible individuals from the disease spreading process.
The majority of epidemiological models described above assume that the disease spreads inside a unique population, e.g., city, region, country. Metapopulation models~\cite{Chinazzi395}
go beyond compartmental models by adding a further spatial dimension and considering a network of spatially separated subpopulations among which individuals can move freely, and come in contact with each other.

Most of these compartmental models describe the flow dynamics from one compartment to another by means of a set of stochastic differential equations. In most cases, the main model parameters are fixed and do not vary with time.
In our proposed framework, described in the following sections, we assume that relevant epidemiological model parameters are time-varying to better capture the effects of mobility and possible restrictive measures. These parameters are then estimated online along with the
epidemiological model states.

\vspace{-3mm}
\section{Learning, detection and forecast}
\label{sec:proposed_method}

Our proposed decision-directed estimation (learning)-detection-forecasting framework 
is presented in Fig.~\ref{fig:flowchart}.
The sketch reads from left to right, and describes the main stages using
the SIR epidemiological model; nevertheless, other models, as those introduced in Section \ref{sec:epidemic_modeling}, can be employed.
Moreover, we describe the framework using the sequences of daily new positive individuals and the cumulative number of healed people and fatalities, that are grouped under the ``recovered'' or removed compartment.
The use of these sequences, among many others, is due to the fact that this information is available for a large number of countries and territories,
which makes the proposed framework directly applicable to data from different areas of the world.
Nevertheless, the algorithm is general enough to be extended and used with different and richer data including, but not limited to,
the number of swab tests and the sequence of hospitalized individuals~\cite{braca2020quickest_1}.
In this regard, we note that the use of the sequence of hospitalized individuals for the quickest detection of a pandemic wave has been recently investigated in \cite{eusipco21}.
The analysis shows that, even though the number of hospitalized individuals is less susceptible to reporting errors, the detection obtained by using the number of infected individuals is usually quicker.
The following subsections provide a \textit{tutorial} description of each stage of the framework.

\vspace{-2mm}

\subsection{Bayesian learning of pandemic evolution}
\label{subsec:adaptive_learning}
The objective of the \textit{Bayesian learning} step, shown at the
top of Fig. \ref{fig:flowchart}, is to track the day-by-day evolution of the epidemiological model states, specifically the number of infected (I) and recovered (R) individuals,
as well as the model parameters, i.e., infection rate $\beta$ and recovery rate $\gamma$, through the daily (possibly partial) observations of the states.
These observations are affected by a certain level of randomness, unavoidable in real-world measurements, modeled as superimposed ``noise''.

The approach we adopt,
recently proposed in~\cite{GagBraMilSolForMar:J20}, is based on the 
discretization of the continuous stochastic differential equations 
that describe the compartmental epidemiological model, and on the assumption that
the model 
parameters
are
time-varying.
Then,
by applying basic principles of Bayesian sequential estimation, that involve a prediction and an update step, the posterior probability density functions (pdfs) of the model parameters, as well as the pdfs of the model states, are computed.
Specifically, during the prediction step,
the
pdfs of the
parameters $\beta$ and $\gamma$, and of the
states I and R, obtained the day before, are predicted
according to the dynamic model defined by the set of discrete-time stochastic difference equations;
during the update step, the new observations
are processed and used to
refine those predicted pdfs, finally providing
the posterior pdfs of model states and parameters at the current time.
The pictorial graphs within the \textit{Bayesian learning} box in Fig.~\ref{fig:flowchart} are examples of the estimated infection and recovery rates over time, and of their confidence intervals.
An efficient implementation of the proposed method, based on mixture models, is presented in~\cite{GagBraMilSolForMar:J20}; therein,
a concrete example of the application to Italian and US data
is also provided. The same implementation, enhanced by the information provided by the \textit{quickest detection} step, will be used  for the data analysis in Section~\ref{sec:analysis_second_wave}.

\vspace{-4mm}

\subsection{Quickest detection of pandemic onset}
\label{subsec:quick_det}

The \textit{quickest detection} step, shown at the bottom of Fig. \ref{fig:flowchart}, is designed to recognize, as quickly as possible, the passages from a controlled to a critical regime of the pandemic, and vice-versa.
The detection procedure that we adopt, proposed in \cite{braca2020quickest_1,braca2020quickest_2}, is based on the growth rate sequence, computed daily as the ratio between two consecutive new positive case counts; this is preceded by a pre-processing of the sequence of daily new positive individuals  to mitigate gross errors and weekly fluctuations in the reported data.
Intuitively, if the growth rate of infected individuals is below unity, the pandemic is under control and will  wane; otherwise, if the growth rate is above $1$, the contagion is still spreading.
However, the growth rate is randomly fluctuating, and its simple observation is not adequate to declare the onset or termination of a pandemic wave, thus requiring the design of a specific statistical test.

The growth rate is modeled as normally distributed with unknown and time-varying mean, whereas the standard deviation is re-estimated daily from a sliding window of the available data.
To detect regime transitions, we rely on the generalized likelihood ratio test (GLRT) approach, see e.g.~\cite{braca2020quickest_2},
which has proven its effectiveness in applications with unknown parameters in the statistical distribution of data.
Specifically, the GLRT solution to the quickest detection problem of interest amounts to 
recursively computing 
the 
MAST decision statistic
that depends only on the observed growth rate and on its estimated standard deviation.
{Then, a regime change is declared when the MAST decision statistic exceeds a predefined threshold,
which is selected to trade-off the \textit{decision delay}, i.e., 
the average time elapsed from the actual change of regime to the detection,
and the so-called \textit{risk}, i.e.,
the reciprocal of the mean time between two consecutive false alarms;}
a false alarm is defined as a threshold crossing {when no change has occurred}. The risk plays the role of the \textit{false alarm probability} in the classical detection theory.  
The results on the performance of MAST,  
presented in~\cite{braca2020quickest_1,braca2020quickest_2},
show that the decision delay
required to reveal the onset of an exponential phase is in the order of a few days, with a risk that scales exponentially with the delay.
The pictorial graph in the \textit{quickest detection} box of Fig.~\ref{fig:flowchart} shows three MAST statistics exceeding a fixed threshold (dashed horizontal line), each corresponding to, respectively, the onset of the second pandemic wave (continuous magenta line), the termination of the second wave (continuous yellow line), and  the onset of the third wave (dashed magenta line).
A fourth MAST statistic (yellow dashed line) corresponding to the termination of the third wave has   not as of this writing exceeded the threshold; therefore, in the example depicted in Fig. \ref{fig:flowchart}, the pandemic is still in a critical regime.
As described in the next subsection, the output of the \textit{quickest detection} step controls the forecast of the pandemic evolution.

\vspace{-5mm}
\subsection{Forecasting of pandemic evolution}
\label{subsec:forecasting_epidem_evol}

Once a transition from a controlled to a critical regime, or vice-versa, is detected through MAST, an infection rate evolution strategy
is hypothesized. An infection rate strategy is the
hypothesized evolution of infection rate $\beta$ that depends on its natural evolution and how the authorities and population respond to regime transitions.
We call them "scenarios".
Specifically, when an outbreak is declared (critical regime), the hypothesized infection rate slope (i.e., the derivative of the infection rate continuously estimated through the Bayesian learning algorithm) is positive (or zero) as shown in the lowermost pictorial graph in the \textit{parameters forecast} box of Fig.~\ref{fig:flowchart}; whereas when the termination of a pandemic wave is declared (controlled regime), the hypothesized infection rate slope is negative (or zero) as shown in the uppermost pictorial graph.
Then, due to the fact that the SIR epidemiological model is nonlinear (as are all the epidemiological models described in Section II), the forecast of the pandemic evolution is produced via ensemble forecasting (see \cite{GagBraMilSolForMar:J20} and references therein), i.e., a Monte Carlo approach that produces a set (or ensemble) of forecasts.
Specifically, first the posterior pdf of the epidemiological model states (I and R) is sampled to obtain an initial ensemble, then this ensemble is propagated forward in time --- according to the epidemiological dynamic model and using the hypothesised infection rate --- up to a forecast horizon of $K$ days.
Evidently, the quality of the
pandemic forecasts depends on the assumed evolution of the infection rate, which, in turn, depends on how authorities and people are expected to respond.
The mean and standard deviation of the ensemble represent, respectively, the evolution of the pandemic, in terms of infected and recovered individuals, and its confidence interval.
The pictorial graphs in the \textit{pandemic forecast} box of Fig.~\ref{fig:flowchart} show the pandemic forecasts in case a critical regime is declared (lowermost graph), and in case a controlled regime is declared (uppermost graph).

\vspace{-2mm}

\section{Analysis of second and third \\ waves in USA}
\label{sec:analysis_second_wave}

\begin{figure}[!t]
\centering

\psfragfig[width=.45\textwidth, height = 4.6cm]{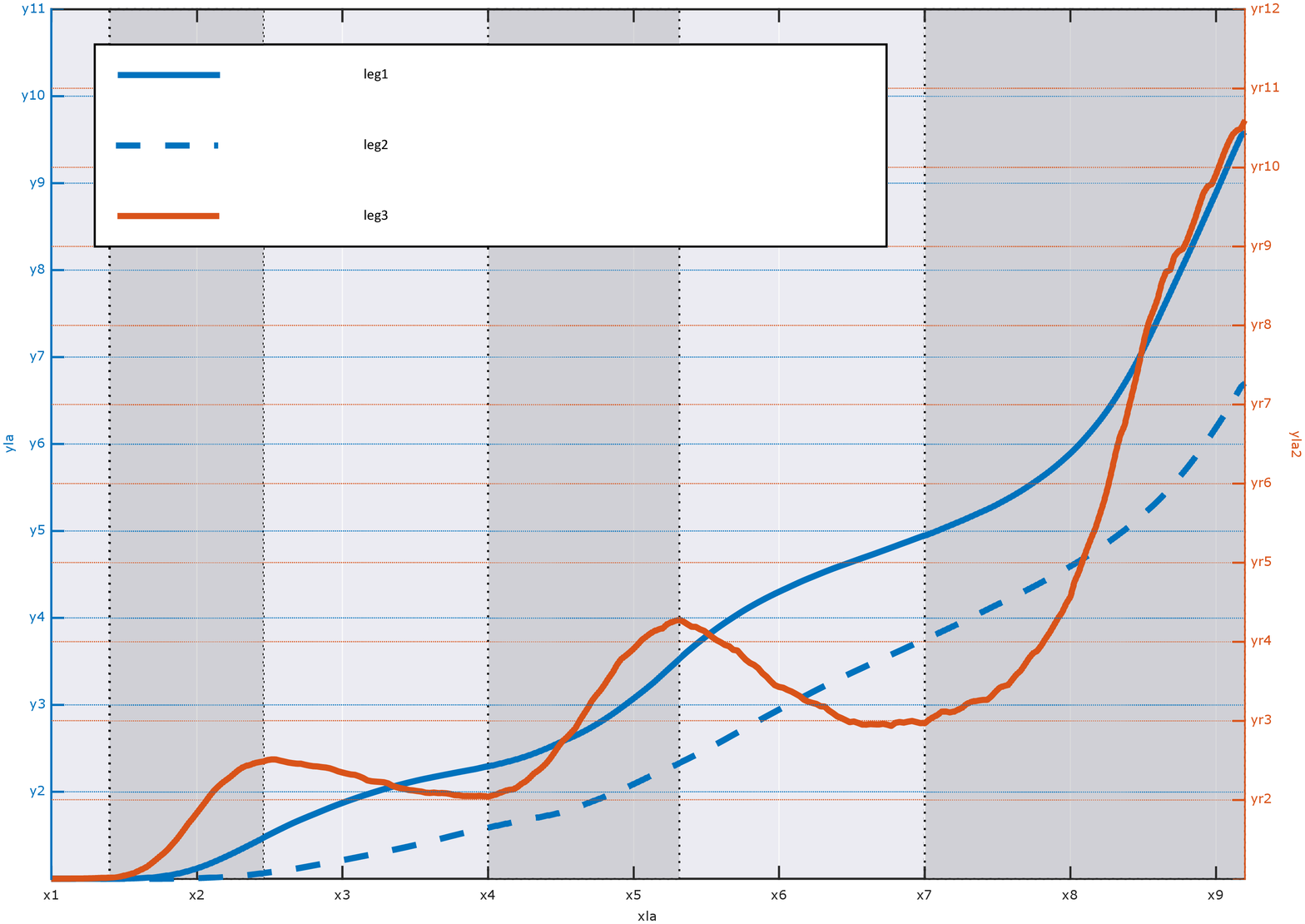}
\vspace{2mm}
\caption{21-day average of numbers of infected and recovered individuals (left axis), and number of daily new positive individuals (right axis) in the USA since March 1, 2020 (data from JHU \cite{DonDuGar:J20}).
Darker areas correspond to the three waves of the
pandemic, characterized by exponential growth in the number of daily new positives.}
\vspace{5mm}
\label{fig:data}

\psfragfig[width=.43\textwidth, height = 4.6cm]{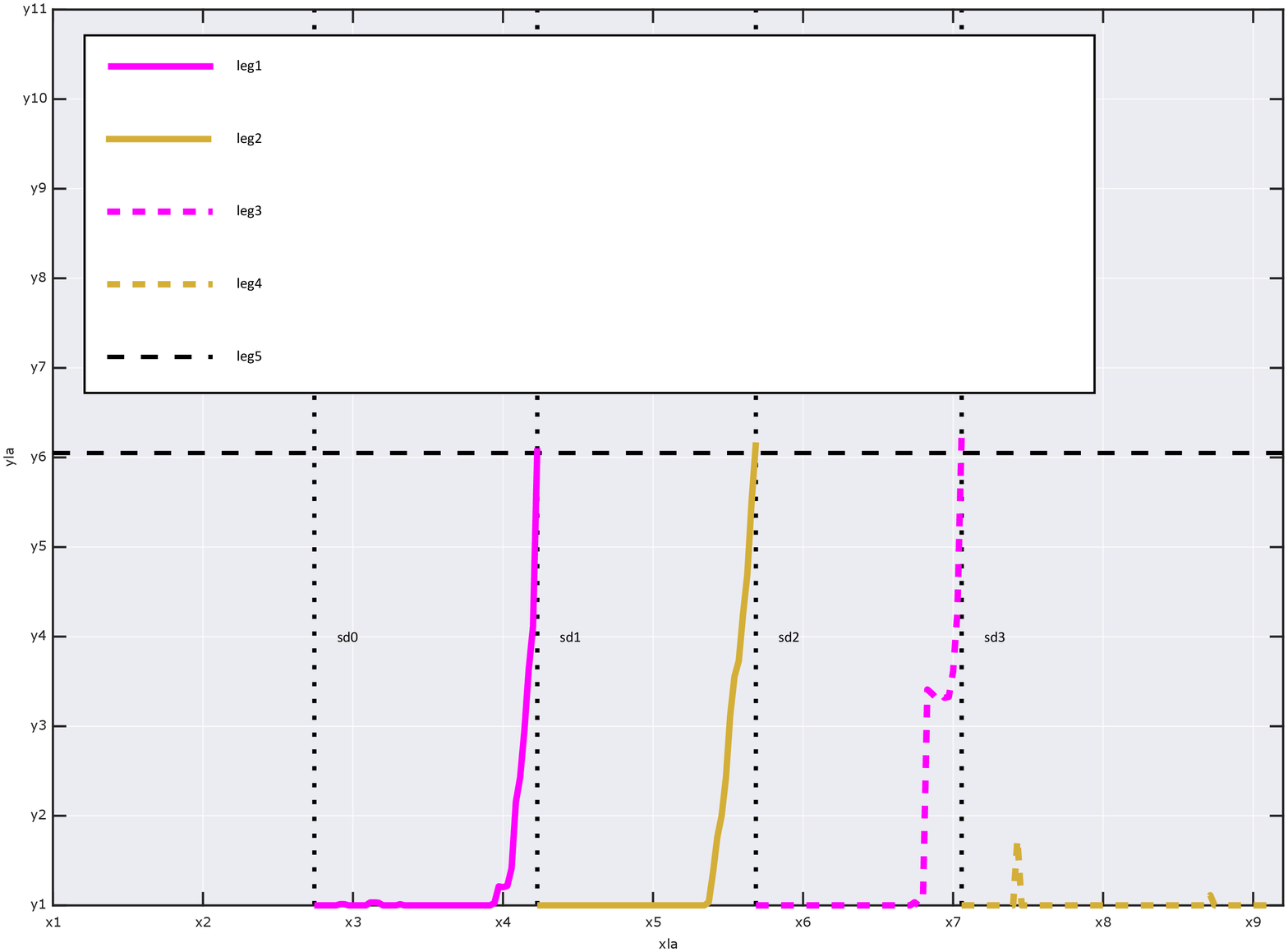}
\vspace{2mm}
\caption{MAST statistics are computed, starting from May 1,
for the onset detection 
of the second wave (magenta solid line) and the third wave (magenta dashed line), and for the termination detection of the second wave (dark yellow solid line) and the third wave (dark yellow dashed line).
The threshold (black dashed line)
corresponds to a risk level of $10^{-4}$.
The onset of the second wave is declared on June 22, and its termination on August 12.
The onset of the third wave is declared on September 29, and it is still ongoing.
}
\vspace{-6mm}
\label{fig:mast}

\end{figure}

The proposed
decision-directed forecasting framework is applied to the COVID-19 dataset from the USA in order to recognize
{the beginning and the termination of the second and
third waves, and infer the progression of the pandemic.}
As described earlier, Bayesian learning uses the number of infected individuals and the number of recovered individuals (which encompasses
the total recoveries plus deaths), while the quickest detection, implemented through the MAST, uses the number of daily new positive individuals.
These numbers, along with many others, are provided daily by the authorities, and have been  collected and made publicly available by the Johns Hopkins University (JHU) since the beginning of the COVID-19 emergency \cite{DonDuGar:J20}.
Fig. \ref{fig:data} shows the 21-day average of numbers of infected, recovered, and daily new positives from March 1
to December 13, 2020.
There are clearly three identifiable waves, each characterized by rapid growth in the number of daily new positives; these regions of exponential growth are highlighted with a darker background for clarity. {Actually, the third wave is likely to be a delayed second wave in different geographic regions of the USA, as a state-by-state analysis seems to imply.}

The MAST statistic used for the onset detection of the second wave is calculated from May 1, and  is  shown in {solid} magenta in Fig. \ref{fig:mast}.
The threshold is {obtained from the analysis in \cite{braca2020quickest_1}, and} is set assuming a risk level of $10^{-4}$, which corresponds to accepting, on average, a false detection every 27 years.
Note that different risk levels might be used for the detection of the onset and the termination of a wave; however, previous analyses have shown that this would change the time a detection is declared by only a few days \cite{eusipco21}.
The onset of the second wave is declared on June 22,
the day on which the statistic crosses the threshold for the first time.
On this day, the MAST statistic  
to detect the termination of the second wave is also initiated, shown in solid dark yellow in Fig. \ref{fig:mast}, leading to a termination detection on August 12.
Likewise, the onset of the third wave is declared on September 29, and it is still ongoing as of December 13.

\begin{figure}[!t]
\centering

\psfragfig[width=.43\textwidth, height = 5.1cm]{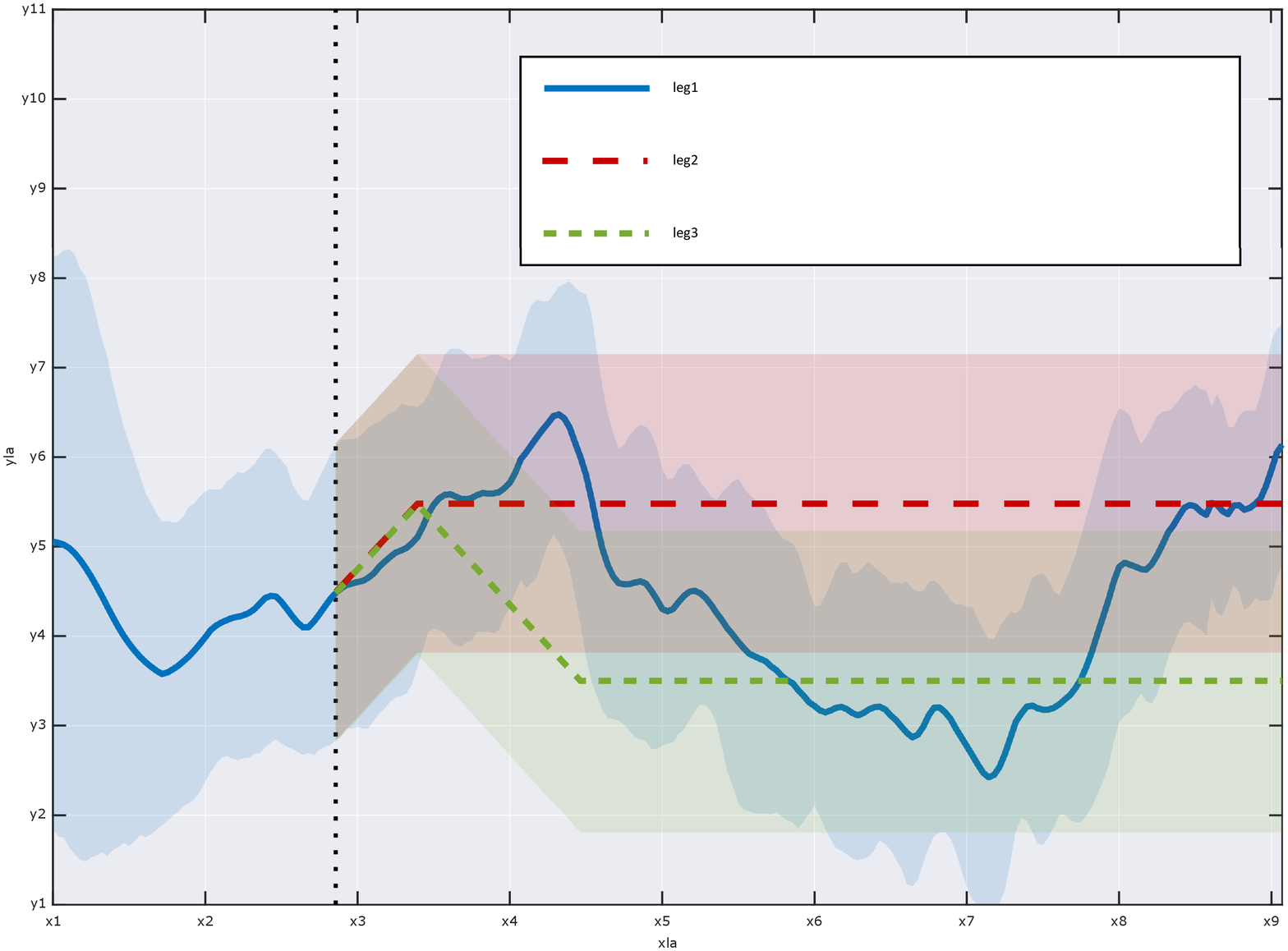}	\vspace{2mm}
\caption{The blue solid line represents the infection rate estimated from May 1 to December 13; the light blue area is its 90\% confidence interval.
Forecast scenario A (red dashed line) and forecast scenario B (green dashed line) are two possible progressions of the infection rate envisaged on June 22, the day on which the onset of the second wave is detected; the light red area and the light green area represent the 90\% confidence interval of the two forecasts, respectively.}
\vspace{6mm}
\label{fig:beta}

\psfragfig[width=.43\textwidth, height = 5.1cm]{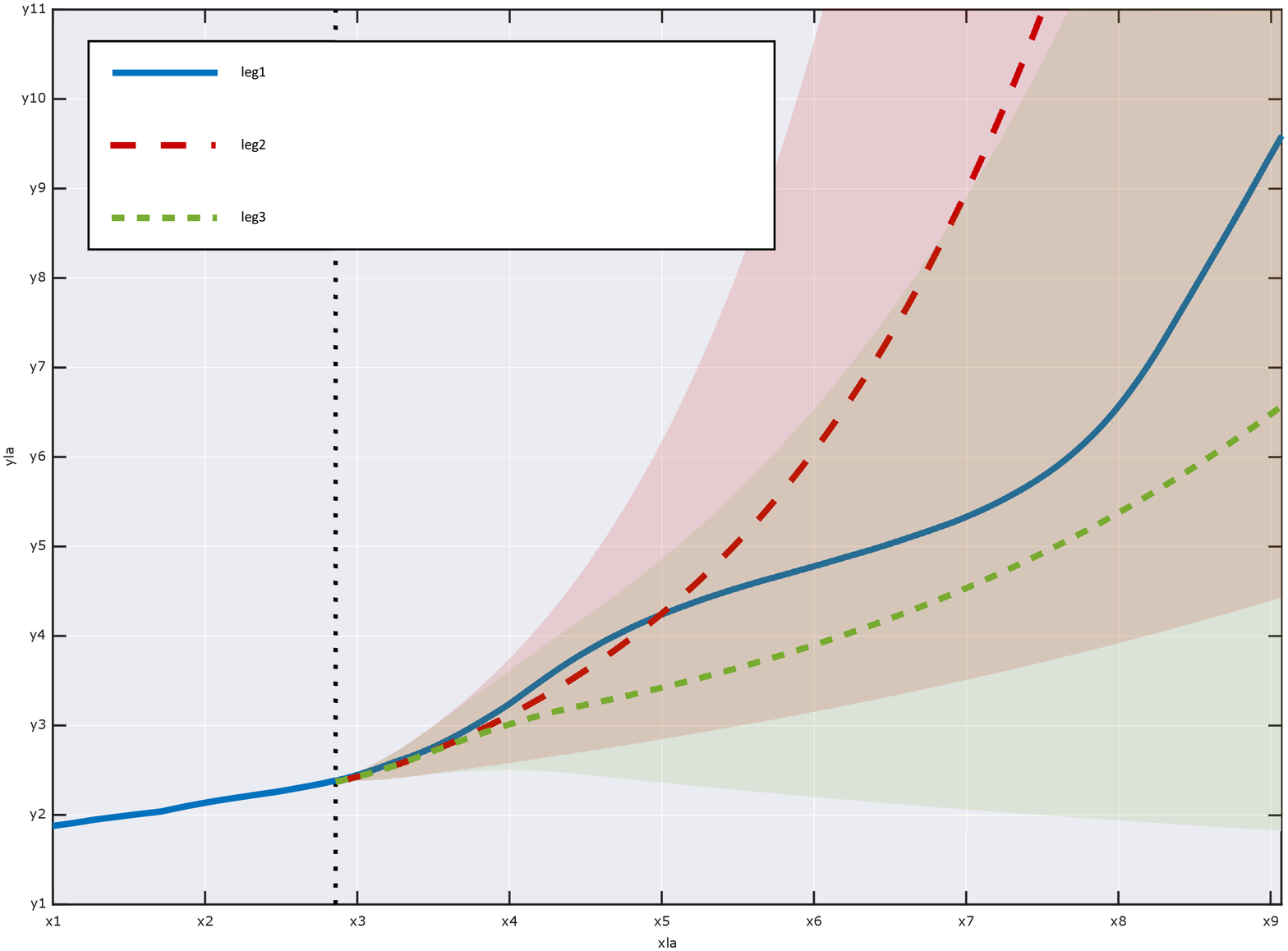} 	\vspace{2mm}
\caption{Evolution of the number of infected individuals according to forecast scenario A (red dashed line) and forecast scenario B (green dashed line) from June 22, the day on which the onset of the second wave is detected. The blue solid line represents the observed number of infected individuals from May 1 to December 13.}
\vspace{6mm}
\label{fig:infected}

\psfragfig[width=.43\textwidth, height = 5.1cm]{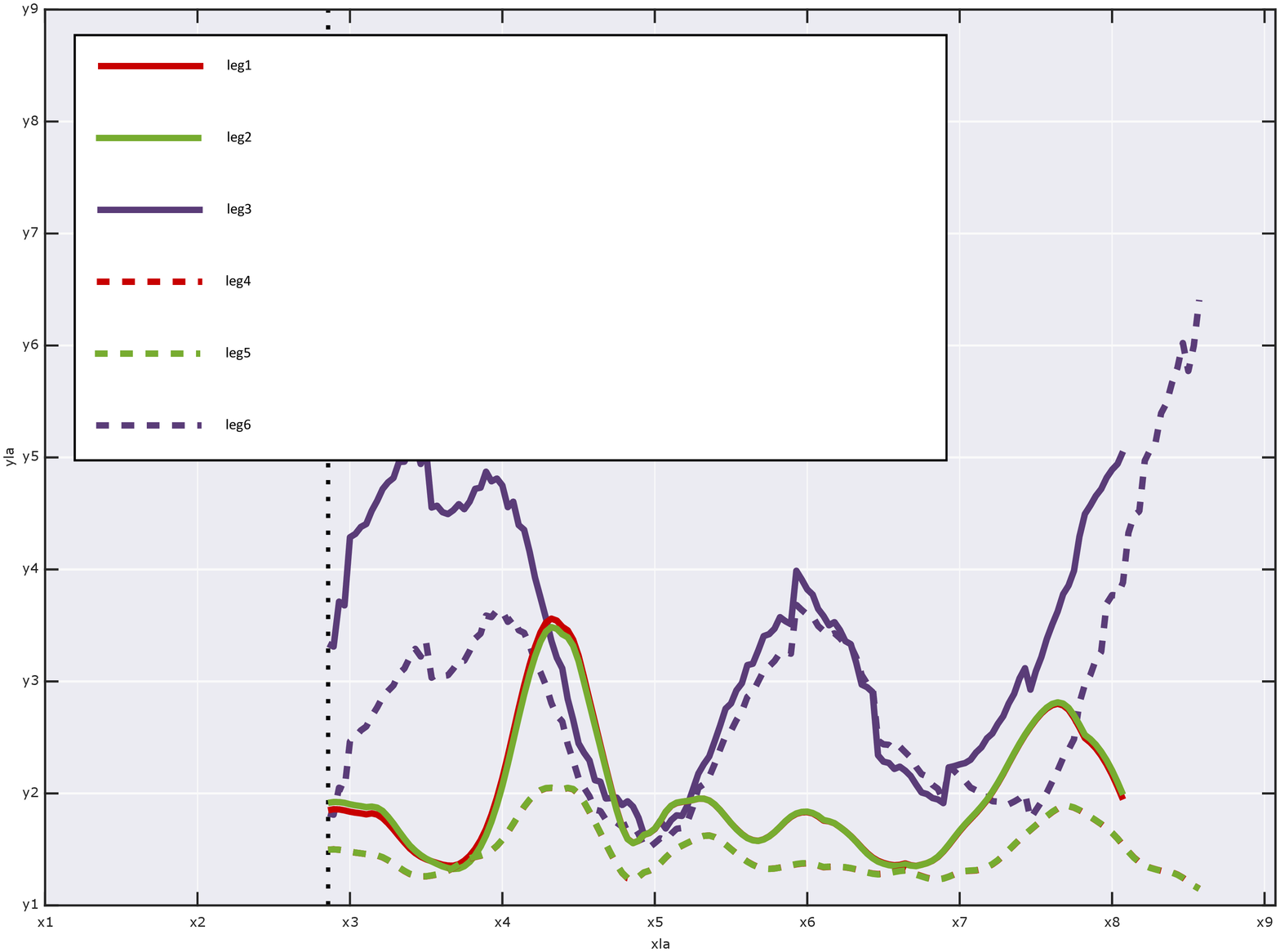} \vspace{2mm}
\caption{Mean absolute percentage error (MAPE) of the forecast of the
pandemic evolution performed with the proposed algorithm (for scenarios A and B) and the GSEIR-fit, on different days (abscissa) and for different time horizons, i.e., 2 and 4
weeks, (depicted with solid and dashed
lines, respectively).
}
\label{fig:performance}

\vspace{-10mm}

\end{figure}

While the MAST statistic is computed for detection purposes, the Bayesian learning algorithm continuously estimates the epidemiological model's states and parameters.
As an example, Fig. \ref{fig:beta} shows in blue the infection rate estimated from May 1 to December 13, along with its 90\% confidence interval.
For forecasting, one needs to incorporate control policies in the form of scenarios. To do this,  from June 22 (day of detection of the second wave) onward, 
two possible progressions of the infection rate are envisaged and the concomitant forecasts  reported.
Forecast ``A'', depicted with a dashed red line, assumes that the infection rate so far learned keeps increasing (or decreasing) with the same slope
for 15 days (this is a reasonable assumption given the range of COVID-19's incubation period), then maintains the attained value for the remaining period; this mimics a scenario in which no countermeasures are taken to slow down the infection rate. Forecast ``B'', instead, mimics a scenario in which restrictions are applied to limit the
pandemic. Therefore, if, on the day of forecasting, the infection rate is increasing, one assumes that it keeps increasing for 15 days with the same estimated slope, then decreases for 30 days with the opposite slope (to model the rightward skew in infection distribution), and finally maintains the attained value for the remaining period; this is illustrated with a dashed green line in Fig. \ref{fig:beta}.
On the other hand, if, on the day of forecasting, the infection rate is decreasing, it keeps decreasing for 15 days and then maintains the realized value for the remaining period, as in forecast scenario A. {Unlike \cite{GagBraMilSolForMar:J20}, the estimated slope of the infection rate on a given day is
computed by averaging the slopes since the day MAST declares a change (either of the onset or the termination of a pandemic wave). If the slope is not coherent with the declared regime (that is, positive slope under critical regime and negative slope under controlled regime) because of the random fluctuation in the infection rate estimation, then the slope used in the forecast is set to zero. Other strategies can be investigated for improving the forecast reliability; however, such investigations are delegated to future work.}

The evolution of the
pandemic in terms of the number of infected individuals in these two cases, i.e., forecast scenario A and forecast scenario B, is shown in Fig. \ref{fig:infected} with a dashed red line and a dashed green line, respectively, and compared with the true number of infected individuals, i.e., the blue solid line.
Both the forecasts
follow the actual evolution of the number of infected, particularly in the
next
1-2 months.
Forecast
B clearly foresees a lower number of infected compared to forecast A, since it is assumed that countermeasures to the spreading of the virus will take place after 15 days from the detection of the onset.
A more rigorous evaluation of the proposed Bayesian learning and forecast algorithm is provided in Fig. \ref{fig:performance}, that shows the mean absolute percentage error (MAPE) computed on the number of infected individuals for both strategies and for
two different forecast horizons $K$, i.e.,
2 and 4 weeks.
Apart from the time intervals of 
roughly between July 19 and August 13, and October 22 and November 14, the MAPE is below 5\% for both forecast scenarios A and B.
Between July 19 and August 13, the MAPE increases --- {still below 5\% and 15\% for horizons of 2 and 4 weeks, respectively} --- as effect of the reduction of the infection rate that follows its peak reached on August 2 (see Fig. \ref{fig:beta}).
Indeed, independently of the forecast scenario, one still assumes --- without any further knowledge --- that the infection rate keeps increasing for 15 days, and the closer one gets to August 2, the more the
hypothesized infection rate deviates from the actual one.
The same happens at the end of October when the infection rate, after attaining its minimum value, starts increasing again.
A comparison with an alternative curve-fitting approach, called GSEIR-fit~\cite{PenYanZhaZhuHon:J20}, is also provided. The GSEIR-fit employs a nonlinear least squares fitting algorithm that, using the number of infected and recovered individuals, computes the six parameters of the GSEIR compartmental model 
and then uses them to forecast the evolution of the
pandemic.
As shown in Fig. \ref{fig:performance}, the proposed forecast algorithm outperforms the GSEIR-fit approach for both forecast horizons of 2 and 4 weeks.
Even for longer-term forecasts (not reported in Fig. \ref{fig:performance}), e.g., 8 weeks, the time-averaged MAPE is 12.2\% and 10.9\% for the proposed algorithm with forecast strategies A and B, respectively, and 16.2\% for the GSEIR-fit approach.
It is worth noting that the same curve-fitting approach using the SIR and SIR-X epidemiological models leads to less accurate forecasts than that obtained with the GSEIR-fit, and, consequently, to those obtained with the proposed algorithm.
Indeed, the forecast obtained with the SIR-fit approach presents a time-average MAPE of 101.8\% and 137.2\% for forecast horizons of 2 and 4 weeks, respectively; the time-averaged MAPE obtained with the SIR-X-fit approach, instead, is 25.9\% and 27.5\% for forecast horizons of 2 and 4 weeks, respectively.
 
\vspace{-3mm}

\section{Conclusion}
\label{sec:conclusion}

Leveraging known concepts from the fields of signal processing and communication,
we have proposed an integrated detection-estimation-forecasting framework that is able to reliably detect the onset and termination of pandemic waves, as well as to 
forecast the epidemiological evolution.
A pandemic wave onset (termination) is determined by an infection rate increase (decrease), also referred to as critical (controlled) regime. The detection of such regimes and the ability to
learn relevant epidemiological factors are crucial to determine an infection rate evolution scenario for reliable
pandemic forecasting.
Experimental validation on COVID-19 data from the USA has shown that the proposed framework is able to reliably detect two consecutive exponential outbreaks on June 22 and September 29, 
and forecast the
pandemic evolution over time horizons ranging from 2 to 4 weeks,
while maintaining a mean absolute percentage error between 5\% to 15\%.

Learning and forecasting, as described in this paper, are based on the classical SIR model. 
However, the proposed methodology is general enough to be able to accommodate more detailed compartmental models.
They would allow the modeling of additional mechanisms, such as the effect of the vaccination campaign and social distancing measures, as well as predict the evolution of other metrics, such as hospitalizations. 
Further extensions
might include the use of metapopulation models to better describe the diffusion of the infection among geographically distributed subpopulations.

\vspace{-3mm}

\renewcommand{\baselinestretch}{.98}

\bibliographystyle{IEEEtran}
\bibliography{IEEEabrv,biblio}
\vskip -2.5\baselineskip plus -1fil
\begin{IEEEbiographynophoto}{Giovanni Soldi}
is a Scientist at
CMRE. He received his Ph.D. degree in Signal Processing from T\'el\'ecom ParisTech in 2016. \end{IEEEbiographynophoto}
\vskip -2.8\baselineskip plus -1fil
\begin{IEEEbiographynophoto}{Nicola Forti}
is a Scientist 
at CMRE. 
He received his Ph.D. degree in Information Engineering from
University of Florence, Italy, in 2016.
\end{IEEEbiographynophoto}
\vskip -2.8\baselineskip plus -1fil
\begin{IEEEbiographynophoto}{Domenico Gaglione} is a Scientist at CMRE.
He obtained the Ph.D. degree
in Signal Processing from University of Strathclyde
in 2017. \end{IEEEbiographynophoto}
\vskip -2.8\baselineskip plus -1fil
\begin{IEEEbiographynophoto}{Paolo Braca}
is a Scientist at CMRE. 
He received the Ph.D. degree in Information Engineering from
University of Salerno, Italy, in 2010.\end{IEEEbiographynophoto}
\vskip -2.8\baselineskip plus -1fil
\begin{IEEEbiographynophoto}{Leonardo M. Millefiori} is a Scientist at CMRE.
He received the M.Sc. degree in Communication
Engineering from
Sapienza University of Rome, Italy, in 2013. \end{IEEEbiographynophoto}
\vskip -2.8\baselineskip plus -1fil
\begin{IEEEbiographynophoto}{Stefano Marano}
received the Ph.D. degree in Electronic Engineering and Computer Science 
from the University of Naples, Italy, in 1997. He
is currently a Professor with DIEM, University of Salerno, Italy.
\end{IEEEbiographynophoto}
\vskip -2.8\baselineskip plus -1fil
\begin{IEEEbiographynophoto}{Peter K. Willett} received the Ph.D. degree from Princeton University, NJ, in 1986.
He is currently a Professor with the Department of Electrical and Computer Engineering, University of Connecticut, Storrs, CT, USA. 
\end{IEEEbiographynophoto}
\vskip -2.8\baselineskip plus -1fil
\begin{IEEEbiographynophoto}{Krishna R. Pattipati} is the Board of Trustees Distinguished Professor and the UTC Chair Professor in systems engineering with the Department of Electrical and Computer Engineering, University of Connecticut, Storrs, CT, USA.
\end{IEEEbiographynophoto}

\end{document}